\documentclass[11pt,epsf,letterpaper]{article}%
\usepackage{setspace}
\usepackage{amsmath}
\usepackage{amsfonts}
\usepackage{verbatim}
\usepackage{amssymb}
\usepackage{graphicx}%
\providecommand{\U}[1]{\protect\rule{.1in}{.1in}}

\begin{document}

\title{A new cubic theory of gravity in five dimensions: Black hole, Birkhoff's
theorem and C-function.}
\author{Julio Oliva$^{1}$ and Sourya Ray$^{2}$\\$^{1}$Instituto de F\'{\i}sica, Facultad de Ciencias, Universidad Austral de
Chile, Valdivia, Chile.\\$^{2}$Centro de Estudios Cient\'{\i}ficos (CECS), Casilla 1469, Valdivia, Chile.\\{\small {CECS-PHY-10/03}}\\{\small {julio.oliva@docentes.uach.cl, ray@cecs.cl}}}
\maketitle

\begin{abstract}
We present a new cubic theory of gravity in five dimensions which has second
order traced field equations, analogous to BHT new massive gravity in three
dimensions. Moreover, for static spherically symmetric spacetimes all the
field equations are of second order, and the theory admits a new
asymptotically locally flat black hole. Furthermore, we prove the uniqueness
of this solution, study its thermodynamical properties, and show the existence
of a C-function for the theory following the arguments of Anber and Kastor
(arXiv:0802.1290 [hep-th]) in pure Lovelock theories. Finally, we include the
Einstein-Gauss-Bonnet and cosmological terms and we find new asymptotically
AdS black holes at the point where the three maximally symmetric solutions of
the theory coincide. These black holes may also possess a Cauchy horizon.

\end{abstract}
\tableofcontents

\section{A cubic theory in five dimensions}

There has been a considerable interest in higher curvature theories of gravity
in the last few decades. Among them, the most prominent one is Lovelock theory
of gravity, which is a natural generalization of Einstein's General Relativity
in higher dimensions \cite{LOVELOCK}. Lovelock theories are characterized by
the special property that the field equations are of at most second order in
derivatives of the metric. As a consequence, they can have stable, ghost-free,
constant curvature vacua \cite{BOULWARE-DESER}. In five dimensions, the most
general Lovelock gravity is given by an arbitrary linear combination of the
cosmological constant, the Ricci scalar curvature and the four-dimensional
Euler density (also known as the Gauss-Bonnet term) which is quadratic in
curvature. The Lovelock terms which are higher order in curvature vanish
identically in five dimensions. However, recently another higher curvature
theory in three dimensions has drawn a lot of attention. This theory, known as
the BHT New Massive Gravity \cite{BHT1}, supplements to the usual
Einstein-Hilbert term, a precise combination of quadratic curvature
invariants. One of the key properties of the theory is that the trace of the
field equations arising from the pure quadratic part, being proportional to
itself, is of second order. The pure quadratic part is the unique quadratic
curvature invariant which possesses this property in three dimensions
\cite{NO}, \cite{far}. In five dimensions, there are two linearly independent
cubic curvature invariants which share this property \cite{cubicwork}. One of
them can be expressed as a complete contraction of three conformal Weyl
tensors. In this section, we present another linearly independent cubic
curvature invariant (in five dimensions), which shares this property with the
pure quadratic BHT (in three dimensions). Consider the following action in
five dimensions%
\begin{equation}
I=\kappa\int\sqrt{-g}\mathcal{L\ }d^{5}x\ , \label{action}%
\end{equation}
where $[\kappa]=[Length]$ is assumed to be positive hereafter, and the
Lagrangian is given by%
\begin{equation}
\mathcal{L}=-\frac{7}{6}R_{\ \ cd}^{ab}R_{\ \ bf}^{ce}R_{\ \ ae}^{df}%
-R_{ab}^{\ \ cd}R_{cd}^{\ \ be}R^{a}_{\ e}-\frac{1}{2}R_{ab}^{\ \ cd}%
R_{\ c}^{a}R_{\ d}^{b}+\frac{1}{3}R_{\ b}^{a}R_{\ c}^{b}R_{\ a}^{c}-\frac
{1}{2}RR_{\ b}^{a}R_{\ a}^{b}+\frac{1}{12}R^{3}\ . \label{lag}%
\end{equation}
Varying the action with respect to the metric gives the following fourth order
field equations
\begin{align}
E_{ab} &  =-\frac{7}{6}\left[  3R_{ahd}^{\ \ \ g}R_{b}^{\ prd}R_{pgr}%
^{\ \ \ h}-3\nabla_{p}\nabla_{q}(R_{\ g\ h}^{p\ q}R_{a\ b}^{\ g\ h}%
-R_{\ hbg}^{p}R_{a}^{\ gqh})-\dfrac{1}{2}g_{ab}R_{\ \ cd}^{mn}R_{\ \ nf}%
^{ce}R_{\ \ me}^{df}\right]  \nonumber\\
&  -\left[  R_{acbd}R^{cspq}R_{pqs}^{\ \ \ d}-R_{a}^{\ qcd}R_{cdb}%
^{\ \ \ h}R_{qh}+R_{b}^{\ dqc}R_{adc}^{\ \ \ h}R_{qh}-\nabla_{p}\nabla
_{q}(R_{ah}R_{b}^{\ phq}+R_{ah}R_{b}^{\ qhp}+R_{bh}R_{a}^{\ phq}\right.
\end{align}
\begin{align}
&  \left.  +R_{\ h}^{q}R_{a\ b}^{\ h\ p}+R_{\ h}^{p}R_{a\ b}^{\ q\ h}%
+\dfrac{1}{2}(g^{pq}R_{a}^{\ hcd}R_{bhcd}+g_{ab}R^{prcd}R_{\ rcd}^{q}%
-g_{a}^{\ p}R_{b}^{\ rcd}R_{\ rcd}^{q}-g_{b}^{\ p}R_{a}^{\ rcd}R_{\ rcd}%
^{q}))\right. \nonumber\\
& \left. -\frac{1}{2}g_{ab}R_{mn}^{\ \ cd}R_{cd}^{\ \ ne}R_{\ e}^{m}\right]-\frac{1}{2}\left[  R_{ac}R_{b}^{\ fcd}R_{fd}+2R_{acbd}R^{cfdg}%
R_{fg}+\nabla_{p}\nabla_{q}(R_{ab}R^{pq}-R_{a}^{\ p}R_{b}^{\ q}+g^{pq}%
R_{acbd}R^{cd}\right.  \nonumber\\
&  \left.  +g_{ab}R^{pcqd}R_{cd}-g_{a}^{\ p}R_{\ cbd}^{q}R^{cd}-g_{b}%
^{\ p}R_{\ cad}^{q}R^{cd})-\dfrac{1}{2}g_{ab}R_{mn}^{\ \ cd}R_{\ c}^{m}%
R_{\ d}^{n}\right]  \nonumber\\
&  +\frac{1}{3}\left[  3R_{acbd}R^{ec}R_{e}^{\ d}+\dfrac{3}{2}\nabla_{p}%
\nabla_{q}(g^{pq}R_{a}^{\ c}R_{bc}+g_{ab}R^{ep}R_{e}^{\ q}-g_{b}^{\ p}%
R^{qc}R_{ac}-g_{a}^{\ p}R^{qc}R_{bc})-\dfrac{1}{2}g_{ab}R_{\ n}^{m}R_{\ c}%
^{n}R_{\ m}^{c}\right]  \nonumber\\
&  -\frac{1}{2}\left[  R_{ab}R^{cd}R_{cd}+2RR^{cd}R_{acbd}+\nabla_{p}%
\nabla_{q}(g_{ab}g^{pq}R^{cd}R_{cd}+g^{pq}RR_{ab}-g_{a}^{\ p}g_{b}^{\ q}%
R^{cd}R_{cd}+g_{ab}RR^{pq}\right.  \nonumber\\
&  \left.  -g_{b}^{\ p}RR_{a}^{\ q}-g_{a}^{\ p}RR_{b}^{\ q})-\dfrac{1}%
{2}g_{ab}RR_{\ n}^{m}R_{\ m}^{n}\right]  +\frac{1}{12}\left[  3R^{2}%
R_{ab}+3\nabla_{p}\nabla_{q}(g_{ab}g^{pq}R^{2}-g_{a}^{\ p}g_{b}^{\ q}%
R^{2})-\dfrac{1}{2}g_{ab}R^{3}\right]  \label{EQM}%
\end{align}

Interestingly, the trace of the field equations, being proportional to the
Lagrangian (\ref{lag}), is of second order. Indeed%
\begin{equation}
E_{\ a}^{a}=\frac{1}{2}\mathcal{L\ }. \label{TR}%
\end{equation}
\qquad

The theory defined by action (\ref{action}) has several other interesting
aspects which we will discuss in the following sections. In section \ref{SSS},
we show that for static spherically symmetric ansatz, the field equations
reduce to second order. We then integrate the field equations to obtain the
most general solution in this family, which, for a certain range of the
integration constant, describes a (topological) black hole. In section
\ref{birkhoffstheorem}, we prove a Birkhoff's theorem for the black hole
solution. In section \ref{entropyfuntion}, we show the existence of a
C-function for the theory using Wald's formula. We then briefly discuss the
thermodynamical properties of the topological black hole in section
\ref{blackholethermodynamics}. In section \ref{nonhomogenouscombination}, we
add an Einstein-Gauss-Bonnet and cosmological term to the cubic Lagrangian and
fix the coupling constants so as to have a unique maximally symmetric vacuum.
At this special point, we obtain a asymptotically AdS black hole solution.
Finally, in section \ref{furthercomments}, we propose a conjecture for the
exitance of arbitrary higher order Lagrangians, sharing the above features and
offer some comments. Appendix A contains the general static spherically
symmetric solution for the non-homogeneous cubic combinations, and in Appendix
B we present the quartic generalization in seven dimensions.

\section{Field equations for static spherically symmetric spacetimes}

\label{SSS} Let us consider the metric%
\begin{equation}
ds^{2}=-f\left(  r\right)  dt^{2}+\frac{dr^{2}}{g\left(  r\right)  }%
+r^{2}d\Sigma_{3}^{2}\ , \label{metric1}%
\end{equation}
where $d\Sigma_{3}$ is the line element of a Euclidean three-dimensional space
of constant curvature $\gamma=\pm1,0$. For $\gamma=+1$, since $\Sigma_{3}$ is
locally isomorphic to $S^{3}$, the metric (\ref{metric1}) possesses spherical
symmetry. In the case $\gamma=-1$, the metric $\Sigma_{3}$ corresponds to an
identification of the hyperbolic space $H_{3}$, while for $\gamma=0$,
$\Sigma_{3}$ is locally flat. Note that the metric (\ref{metric1}) is the most
general static metric which is compatible with the isometries of $\Sigma_{3}$.

The field equations (\ref{EQM}), when evaluated on the metric (\ref{metric1}),
reduce to%
\begin{align}
E_{\ t}^{t}  &  =\frac{1}{2r^{6}}\left(  g-\gamma\right)  ^{2}\left(
2g-3g^{\prime}r-2\gamma\right)  \ ,\label{tts}\\
E_{\ r}^{r}  &  =\frac{1}{2fr^{6}}\left(  g-\gamma\right)  ^{2}\left(
2fg-3rgf^{\prime}-2\gamma f\right)  \ ,\label{rrs}\\
E_{\ j}^{i}  &  =\frac{\gamma-g}{f^{2}r^{6}}\left[  gr^{2}\left(
\gamma-g\right)  f^{\prime2}+ff^{\prime}r\left(  5g^{\prime}rg+4\gamma
g-4g^{2}-\gamma rg^{\prime}\right)  \right. \\
&  \left.  +4f^{2}r\left(  \gamma-g\right)  g^{\prime}-2fgr^{2}\left(
\gamma-g\right)  f^{\prime\prime}+4f^{2}\left(  \gamma-g\right)  ^{2}\right]
\delta_{\ j}^{i} \label{ijs}%
\end{align}
where ($^{\prime}$) denotes derivative with respect to $r$, and the indices
$i,j$ run along the base manifold $\Sigma_{3}$. Then, the field equations for
the metric (\ref{metric1}) reduce to a set of nonlinear second order equations
for the functions $f\left(  r\right)  $ and $g\left(  r\right)  $. Let us
first analyze the nontrivial branch, where $g(r)\neq\gamma$. Solving this
system, it is easy to see that the only nontrivial solution within the static
family (\ref{metric1}) is%
\begin{equation}
ds^{2}=-\left(  cr^{2/3}+\gamma\right)  dt^{2}+\frac{dr^{2}}{cr^{2/3}+\gamma
}+r^{2}d\Sigma_{3}^{2}\ , \label{sol1}%
\end{equation}
where $c$ is an integration constant. This metric is asymptotically locally
flat since $R_{\ \alpha\beta}^{\mu\nu}\rightarrow0$ when $r$ goes to infinity,
and has a curvature singularity at the origin, which can be realized by
evaluating the Ricci scalar%
\begin{equation}
R=-\frac{88c}{9r^{4/3}}\ .
\end{equation}
This singularity is hidden by an event horizon when $c$ is positive and
$\gamma=-1$, in which case it is more convenient to rewrite the metric
(\ref{sol1}) as%
\begin{equation}
ds^{2}=-\left(  \left(  \frac{r}{r_{+}}\right)  ^{2/3}-1\right)  dt^{2}%
+\frac{dr^{2}}{\left(  \frac{r}{r_{+}}\right)  ^{2/3}-1}+r^{2}d\Sigma_{3}%
^{2}\ , \label{BH}%
\end{equation}
where $r_{+}=c^{-3/2}$ is the location of the horizon, whose geometry is given
by $H_{3}/\Gamma$, where $\Gamma$ is a freely acting, discrete subgroup of
$O\left(  3,1\right)  $.

In the next section, we prove a Birkhoff's theorem for the solution when the
staticity condition is removed.

Note that there is another branch of solutions for the system (\ref{tts}%
)-(\ref{ijs}), for which $g(r)=\gamma$ and $f(r)$ is an arbitrary function.
This ``degenerated" behavior is also common to all Lovelock theories
possessing a unique maximally symmetric solution \cite{Charmousis}-\cite{DOT3}.

\section{Birkhoff's theorem}

\label{birkhoffstheorem}

Let us consider the metric
\begin{equation}
ds^{2}=-f\left(  t,r\right)  dt^{2}+\frac{dr^{2}}{g\left(  t,r\right)  }%
+r^{2}d\Sigma_{3}^{2}\ , \label{metricd}%
\end{equation}
where again the manifold $\Sigma_{3}$ is a compact manifold of constant
curvature\footnote{In general, in front of $\Sigma_{3}$ there can be a generic
function $F\left(  t,r\right)  $ which after a gauge fixing, depending on the
norm of its gradient can be chosen to be $r^{2}$, $t^{2}$ or a constant.}
$\gamma=\pm1,0$. For this family of spacetimes, the nonvanishing diagonal
components of the field equations are%
\begin{align}
E_{\ t}^{t}  &  =\frac{1}{2r^{6}}\left(  g-\gamma\right)  ^{2}\left(
2g-3g^{\prime}r-2\gamma\right)  \ ,\label{ttd}\\
E_{\ r}^{r}  &  =\frac{1}{2fr^{6}}\left(  g-\gamma\right)  ^{2}\left(
2fg-3rgf^{\prime}-2\gamma f\right)  \ ,\label{rrd}\\
E_{\ j}^{i}  &  =\frac{\gamma-g}{f^{2}g^{2}r^{6}}\left[  g^{3}r^{2}\left(
\gamma-g\right)  f^{\prime2}+ff^{\prime}g^{2}r\left(  5g^{\prime}rg+4\gamma
g-4g^{2}-\gamma rg^{\prime}\right)  \right. \\
&  \left.  +4f^{2}g^{2}r\left(  \gamma-g\right)  g^{\prime}-2fg^{3}%
r^{2}\left(  \gamma-g\right)  f^{\prime\prime}+4f^{2}g^{2}\left(
\gamma-g\right)  ^{2}\right. \\
&  \left.  +gr^{2}\left(  \gamma-g\right)  \dot{f}\dot{g}+fr^{2}\left(
3\gamma+g\right)  \dot{g}^{2}-2fgr^{2}\left(  \gamma-g\right)  \ddot
{g}\right]  \delta_{\ j}^{i} \label{ijd}%
\end{align}
where ($^{\prime}$) and ($\cdot$) denote partial derivatives with respect to
$r$ and $t$ respectively. The off-diagonal component $E_{\ r}^{t}$ contains
mixed partial derivatives of the metric functions. Note that equations
(\ref{ttd}) and (\ref{rrd}) have the same expression as their static
counterpart (\ref{tts}) and (\ref{rrs}) respectively, since they do not
contain derivatives with respect to time.

Equation (\ref{ttd}), is solved by%
\begin{equation}
g\left(  t,r\right)  =F\left(  t\right)  r^{2/3}+\gamma\ ,
\end{equation}
where $F\left(  t\right)  $ is an arbitrary integration function\footnote{When
$F(t)$ is identically vanishing, the remaining field equations are
automatically solved for an arbitrary $f(t,r)=f(r)$, and we get back the
degenerate solution mentioned previously.}. Inserting this expression for
$g\left(  t,r\right)  $ in (\ref{rrd}) we obtain the following equation for
$f\left(  t,r\right)  $:%
\begin{equation}
3r\left(  \gamma+F\left(  t\right)  r^{2/3}\right)  \frac{\partial f\left(
t,r\right)  }{\partial r}-2f\left(  t,r\right)  F\left(  t\right)
r^{2/3}=0\ ,
\end{equation}
whose solution is%
\begin{equation}
f\left(  t,r\right)  =H\left(  t\right)  g\left(  t,r\right)  \ ,
\end{equation}
where $H\left(  t\right)  $ being an integration function, can be absorbed by
a time rescaling. Thus, without any loss of generality, the equations
$E_{t}^{t}=0$ and $E_{r}^{r}=0$ are solved by%
\begin{equation}
f\left(  t,r\right)  =g\left(  t,r\right)  =F\left(  t\right)  r^{2/3}%
+\gamma\ . \label{a=b}%
\end{equation}
Now, after inserting (\ref{a=b}) in (\ref{ijd}), we obtain the following
equation for $F\left(  t\right)  $:%
\begin{equation}
F^{2}\ddot{F}r^{2/3}+\left(  2\dot{F}^{2}+F\ddot{F}\right)  \gamma=0\ .
\label{eijrep}%
\end{equation}
This implies that , for $\gamma\neq0$, $F\left(  t\right)  $ must be a
constant $c$, and the metric (\ref{metricd}) reduces to (\ref{sol1}), which is
the static metric obtained previously. Thus, for $\gamma\neq0$ we have proved
the Birkhoff's theorem, since we have shown that the most general solution of
the theory, within the family of spacetimes (\ref{metricd}), is static and is
given by (\ref{sol1}). For $\gamma=0$, equation (\ref{eijrep}) implies that
$F\left(  t\right)  =et+c$ ($e$ and $c$ being integration constants). However,
the off-diagonal equation $E_{r}^{t}=0$ implies that $e=0$. Thus, we have
proved the Birkhoff's theorem for planar transverse section, i.e. for
$\gamma=0$, as well. It is quite remarkable that, even though the field
equations (\ref{EQM}), are very complex in general, it admits a Birkhoff's theorem!

\section{Entropy function}

\label{entropyfuntion}

The C-function was first introduced by \cite{Zam}, in the context of QFT's in
$1+1$ dimensions, who showed that under the RG flow to lower energies, the
C-function is a monotonically increasing function of the couplings of the
theory. At the fixed points of the flow, the C-function reaches an extremum
and equals the central charge of the Virasoro algebra corresponding to the
infinite dimensional group of conformal transformation in two dimensions.
Later, Sahakian \cite{sa} gave a covariant geometric expression for the
C-function for theories which admit an holographic description.

In \cite{golds}, Goldstein et al, gave a simple expression for the C-function
for static, asymptotically flat solutions of Einstein's gravity in four
dimensions. They showed that when coupled to matter fields satisfying null
energy condition, this function is a monotonically increasing function of the
radial coordinate and coincides with the entropy when evaluated at the
horizon. This work was generalized in the context of AdS/CFT in \cite{dimitru}%
. Recently, it was shown in \cite{KASTORC}, that C-functions also exists for
static spherically symmetric, asymptotically flat spacetimes in Lovelock
gravity. Moreover, the authors showed that there is a non-uniqueness in the
C-function for second or higher order Lovelock theory. They have further shown
the existence of two possible C-functions, provided the matter field satisfies
appropriate energy conditions. Here, following the same lines of argument as
in \cite{KASTORC}, we show that one such C-function also exists for the theory
defined\footnote{In a very recent paper \cite{sinha}, the author constructed a
cubic generalization of the BHT new massive gravity in three dimensions, by
demanding the existence of a C-function.} by (\ref{action}). This is evident
since the field equations for a spherically symmetric spacetime in our theory
have the same functional form as that of a generic pure Lovelock theory
\cite{KASTORC}. However, in five dimensions, the cubic Lovelock Lagrangian,
being identically vanishing, does not give any field equations. Nevertheless,
the theory defined by (\ref{action}) mimics the cubic Lovelock theory for
spherically symmetric spacetimes.

Consider a metric of the form%
\begin{equation}
ds^{2}=-a^{2}\left(  r\right)  dt^{2}+\frac{dr^{2}}{a^{2}\left(  r\right)
}+b^{2}\left(  r\right)  d\Sigma_{3}^{2}\ , \label{metricKastor}%
\end{equation}
where $\Sigma_{3}$ is a Euclidean space of constant curvature $\gamma=\pm1,0$.
For $\gamma=1$, the spacetime (\ref{metricKastor}) has a spherical symmetry.
The relevant components of the field equations (\ref{EQM}) for this metric
read%
\begin{align}
E_{\ t}^{t}  &  =-\frac{1}{b^{6}}\left(  \gamma-a^{2}b^{\prime2}\right)
^{2}\left(  \gamma+3a^{2}bb^{\prime\prime}+3abb^{\prime}a^{\prime}%
-a^{2}b^{\prime2}\right) \label{ttkastor}\\
E_{\ r}^{r}  &  =-\frac{1}{b^{6}}\left(  \gamma-a^{2}b^{\prime2}\right)
^{2}\left(  \gamma+3abb^{\prime}a^{\prime}-a^{2}b^{\prime2}\right)
\label{rrkastor}%
\end{align}
Suppose the theory is coupled to a matter field, which satisfies the
null-energy condition
\begin{equation}
T_{ab}\xi^{a}\xi^{b}\geq0
\end{equation}
for all null-vectors $\xi^{a}$. This implies the following inequality:%
\begin{equation}
E_{\ t}^{t}-E_{\ r}^{r}=-\frac{3a^{2}}{b^{5}}(\gamma-a^{2}b^{\prime2}%
)^{2}b^{\prime\prime}=T_{\ t}^{t}-T_{\ r}^{r}\geq0. \label{nullcondition}%
\end{equation}
Now, if the metric (\ref{metricKastor}) describes a black hole, then due to
cosmic censorship, $b(r)\neq0$ on or outside the horizon $r=r_{+}$. Without
loss of generality, we can the assume $b(r)>0$ on the horizon. For an
asymptotically flat black hole, as $r\rightarrow\infty$, $b(r)\rightarrow\pm
r$. First consider the case $b(r)\rightarrow-r$ as $r\rightarrow\infty$. Since
$b(r)$ is assumed to be positive on the horizon, $b(r_{0})=0$ for some
$r_{+}<r_{0}<+\infty$, which is discarded by cosmic censorship. Hence,
$b(r)\rightarrow r$ as $r\rightarrow\infty$. Now, if $b(r)$ is not a monotonic
function of $r$, then there must exist at least one local minima, i.e.,
$b^{\prime}\left(  r_{c}\right)  =0$ with $b^{\prime\prime}\left(
r_{c}\right)  <0$, for $r_{+}<r_{c}<+\infty$. However, this is ruled out by
(\ref{nullcondition}) since $a^{2}$ is positive outside the horizon. Thus the
monotonicity of $b$ is proved for the theory coupled to matter fields
satisfying the null-energy condition.

We now compute the entropy of a static black hole of the form
(\ref{metricKastor}), using Wald's formula \cite{Wald},\cite{JM}, which is
given by%
\begin{equation}
S=-2\pi\kappa\int_{\Sigma_{3}}\frac{\partial\mathcal{L}}{\partial R_{abcd}%
}\epsilon_{ab}\epsilon_{cd}\hat{\epsilon}\ ,
\end{equation}
where $\epsilon_{ab}$ is the binormal to the horizon cross-section and
$\hat{\epsilon}$ is the volume form induced on the spatial cross section
$\Sigma_{3}$ of the horizon at $r=r_{+}$. Using the Lagrangian (\ref{lag}), we
compute the curvature components and we obtain%
\begin{align}
S  &  =12\pi\kappa\left.  \frac{(\gamma-a^{2}b^{\prime2})^{2}}{b}\right\vert
_{r=r_{+}}Vol\left(  \Sigma_{3}\right)  \ ,\\
&  =12\pi\kappa\frac{\gamma^{2}}{b(r_{+})}Vol\left(  \Sigma_{3}\right)
\ \ \ \ (\because a=0\ \ \text{at}\ r=r_{+}). \label{aldevaluated}%
\end{align}
Since the C-function is a function of the radial coordinate $r$, which matches
the entropy of the black hole when evaluated on the horizon $r=r_{+}$, one can
extend the entropy formula for arbitrary $r$. The C-function is then given by
\begin{equation}
C(r)=12\pi\kappa\frac{(\gamma-{a(r)}^{2}{b^{\prime}(r)}^{2})^{2}}{b}Vol\left(
\Sigma_{3}\right)  \label{SECOND-C}%
\end{equation}
Let us now check the monotonicity of (\ref{SECOND-C}) as a function of outward
radial coordinate, following along the lines of Ref. \cite{KASTORC}. First
note that using the field equation (\ref{ttkastor}), one can write $tt$
component of the stress energy tensor as
\begin{equation}
T_{tt}=\frac{a^{2}}{b^{6}}\kappa\left[  X^{2}(X-\frac{3bX^{\prime}}%
{2b^{\prime}})\right]  \ ,
\end{equation}
where $X:=\gamma-a^{2}b^{\prime2}$. Now, differentiating $C(r)$ we obtain,
\begin{align}
C^{\prime}(r)  &  =\frac{12\pi}{b^{2}}\kappa(2bXX^{\prime}-X^{2}b^{\prime
})Vol\left(  \Sigma_{3}\right) \nonumber\\
&  =\frac{12\pi}{3b^{2}}\kappa\left(  b^{\prime}X^{2}-4b^{\prime}X\left(
X-\frac{3bX^{\prime}}{2b^{\prime}}\right)  \right)  Vol\left(  \Sigma
_{3}\right) \nonumber\\
&  =\frac{4b^{\prime}\pi}{b^{2}}\kappa\left(  X^{2}-\frac{4b^{6}T_{tt}}{\kappa
Xa^{2}}\right)  Vol\left(  \Sigma_{3}\right)  . \label{esta}%
\end{align}
Using null energy condition, we had shown that $b^{\prime}>0$ (See the
paragraph below Eq. (\ref{nullcondition})) and since weak energy condition
implies $T_{tt}>0$, then for both $\gamma=0,-1$, $X$ $<0$, and hence from Eq.
(\ref{esta}), we see that $C^{\prime}>0$. This proves that, for $\gamma=0,-1$,
the function $C(r)$ is a monotonically increasing function. However, our
analysis is inconclusive for $\gamma=1$.

\section{Black hole thermodynamics}

\label{blackholethermodynamics}

In this section, we explore the thermodynamics of the black hole (\ref{BH}),
which as proved in the previous sections, is the unique solution for the
theory (\ref{action}) within the family of spacetimes (\ref{metricd}).

The temperature of the black hole (\ref{BH}) is given by%
\begin{align}
T  &  =\frac{1}{4\pi}\left(  \left(  \frac{r}{r_{+}}\right)  ^{2/3}-1\right)
_{r=r_{+}}^{\prime}\nonumber\\
&  =\frac{1}{6\pi}\frac{1}{r_{+}}\ ,
\end{align}
which is also the case for spherically symmetric black holes in pure Lovelock
theories for arbitrary order $k<\frac{d-1}{2}$\thinspace. Using Wald's entropy
(\ref{aldevaluated}) one obtains%
\begin{equation}
S=\frac{12\pi}{r_{+}}\kappa Vol\left(  \Sigma_{3}\right)  \ .
\end{equation}
And asuming the validity of the first law $dM=TdS$, one finds that the mass of
the black holes is given by\footnote{It would be interesting to compute the
mass by more standard methods like the Hamiltonian analysis.}%
\begin{equation}
M=\frac{Vol\left(  \Sigma_{3}\right)  \kappa}{r_{+}^{2}}\ ,
\end{equation}
where we have fixed the integration constant $M_{0}$ in such a way that when
$r_{+}\rightarrow\infty$ (i.e. flat space) the mass vanishes.

Since the mass in terms of the temperature is given by%

\begin{equation}
M=36\pi^{2}T^{2}\kappa Vol\left(  \Sigma_{3}\right)  \ ,
\end{equation}
the specific heat $C=dM/dT$ is positive%
\begin{equation}
C=72\pi^{2}\kappa T\ Vol\left(  \Sigma_{3}\right)  \ ,
\end{equation}
which implies that the black hole is thermodynamically locally stable.

\section{Nonhomogeneous combinations: Asymptotically AdS black holes}

\label{nonhomogenouscombination}

In five dimensions the most general Lagrangian giving rise to second order
field equations is given by an arbitrary linear combination of the
Gauss-Bonnet, the Einstein's and cosmological terms. We now look for
nontrivial, spherically symmetric solutions when the cubic Lagrangian
(\ref{lag}) is supplemented by a linear combination of the above terms. We
find a new asymptotically AdS black hole for a particular combination, which
in addition to the event horizon has a Cauchy horizon. The field equations
obtained for the theory considered, are given by%
\begin{equation}
\mathcal{E}_{\mu\nu}:=c_{3}E_{\mu\nu}+c_{2}GB_{\mu\nu}+c_{1}G_{\mu\nu}%
+c_{0}g_{\mu\nu}=0\ , \label{nonhom}%
\end{equation}
where $E_{\mu\nu}$ is defined in Eq. (\ref{EQM}), $G_{\mu\nu}$ is the
Einstein's tensor and the Gauss-Bonnet term is defined by%
\begin{equation}
GB_{\mu\nu}:=2RR_{\mu\nu}-4R_{\mu\rho}R_{\quad\nu}^{\rho}-4R_{\ \rho}^{\delta
}R_{\ \mu\delta\nu}^{\rho}+2R_{\mu\rho\delta\gamma}R_{\nu}^{\quad\rho
\delta\gamma}-\frac{1}{2}g_{\mu\nu}(R_{\rho\delta\gamma\lambda}R^{\rho
\delta\gamma\lambda}-4R_{\rho\delta}R^{\rho\delta}+R^{2})\ .
\end{equation}
Equation (\ref{nonhom}) describes the most general cubic theory in five
dimensions \cite{cubicwork}, whose field equations are of second order, for
static spherically symmetric spacetimes (\ref{metric1}).

The constant curvature solutions of this theory%
\begin{equation}
R_{\ \alpha\beta}^{\mu\nu}=\lambda\ \left(  \delta_{\alpha}^{\mu}\delta
_{\beta}^{\nu}-\delta_{\alpha}^{\nu}\delta_{\beta}^{\mu}\right)  \ ,
\label{constantcurvature}%
\end{equation}
fulfill%
\begin{equation}
2c_{3}\lambda^{3}-12c_{2}\lambda^{2}-6c_{1}\lambda+c_{0}=0\ . \label{lambdas}%
\end{equation}
Generically, there are three constant curvature solutions with different radii
(inverse of different cosmological constants), describing three different
maximally symmetric spacetimes, which corresponds to (A)dS or flat space,
depending on whether $\lambda$ is (negative)positive or zero, respectively. In
analogy with Lovelock theories, it is natural to expect that, the space of the
solutions is enlarged when the three different vacua of the theory degenerate
into one \cite{DOT2}, \cite{DOT3}, \cite{DOTWorm}-\cite{ROT}, which occurs
when%
\begin{equation}
c_{2}=-\frac{c_{1}^{2}}{c_{0}}\text{\ },\text{ and }c_{3}=-4\frac{c_{1}^{3}%
}{c_{0}^{2}}\ . \label{tune}%
\end{equation}
In such a case Eq. (\ref{lambdas}) factorizes as%
\begin{equation}
\frac{\left(  c_{0}-2c_{1}\lambda\right)  ^{3}}{c_{0}^{2}}=0\ .
\end{equation}
Consequently for $c_{0}\neq0$, one obtains%
\begin{equation}
\lambda=\frac{c_{0}}{2c_{1}}\ .
\end{equation}
Assuming for simplicity $f\left(  r\right)  =g\left(  r\right)  $ in
(\ref{metric1}) we integrated the field equations to obtain the following
solution%
\begin{equation}
ds^{2}=-\left(  \frac{r^{2}}{l^{2}}-cr^{2/3}+\gamma\right)  dt^{2}%
+\frac{dr^{2}}{\frac{r^{2}}{l^{2}}-cr^{2/3}+\gamma}+r^{2}d\Sigma_{3}^{2}\ ,
\label{solnonhom}%
\end{equation}
where $c$ is an integration constant, $\gamma$ is the curvature of $\Sigma
_{3}$ and $l^{2}:=-\frac{2c_{1}}{c_{0}}$ is the squared AdS radius which is
assumed to be positive. For $c=0$, the spacetime is locally AdS, and in this
case for $\gamma=-1$, the metric reduces to the massless topological black
hole \cite{topblack}. For $\gamma=1$, the metric (\ref{solnonhom}) is
asymptotically AdS with a slower fall-off as compared with the
Henneaux-Teitelboim asymptotic behavior \cite{HenTai}. This again, is similar
to what occurs in Lovelock theories \cite{BHS},\cite{HIDEKI}. The spacetime
(\ref{solnonhom}) has a curvature singularity at the origin which could be
covered by one or two horizons depending on the values of $c$ and $\gamma$.

For $\gamma=1$, and $-\infty<c<6\left(  \frac{2}{l^{2}}\right)  ^{1/3}$, the
metric (\ref{solnonhom}) describes a naked singularity. In the case
$c=6\left(  \frac{2}{l^{2}}\right)  ^{1/3}$ the spacetime (\ref{solnonhom})
describes an extremal black hole with a degenerate horizon located at
$r_{+}=r_{-}=2l$. In the range $c>6\left(  \frac{2}{l^{2}}\right)  ^{1/3}$ the
metric (\ref{solnonhom}) has an event and a Cauchy horizon, which cover the
timelike singularity at the origin.

For vanishing $\gamma$, the singularity at the origin becomes null, and for
positive $c$ this singularity is hidden by an event horizon located at
$r_{+}=\left(  cl^{2}\right)  ^{3/4}$.

Finally, in the case $\gamma=-1$, there exists an event horizon at $r_{+}$ for
any value of $c$, which covers a spacelike singularity at the origin.

\section{Generalization to arbitrary higher order}

Some aspects of the theory defined here, are common to other, well behaved,
theories of gravity, such as Lovelock theory and BHT new massive
gravity\footnote{Note that for the pure BHT theory there is no Birkhoff's
theorem which is explicit from the existence of gravitational solitons
\cite{Tempo}. This non-uniqueness further allows the existence of a very
interesting Lifshitz black hole \cite{AGH}.} where the trace of the field
equations is of second order. Furthermore, there is an interesting similarity
with the cubic Lovelock theory, which was exploited in the construction of the
C-function. The \textquotedblleft spherically symmetric" solution of the pure
cubic Lovelock theory, is given by \cite{ATZ}%
\begin{equation}
ds^{2}=-\left(  \gamma+\frac{c}{r^{\frac{D-7}{3}}}\right)  dt^{2}+\frac
{dr^{2}}{\gamma+\frac{c}{r^{\frac{D-7}{3}}}}+r^{2}d\Sigma_{\gamma,D-2}^{2}\ ,
\label{lovcub}%
\end{equation}
where $c$ is an integration constant. This is valid for $D>6$. Nevertheless,
if one insists on considering $D=5$ in (\ref{lovcub}), one obtains the metric
(\ref{sol1}). This is also the case for pure BHT new massive gravity
\cite{Tempo},\cite{BHT2}, when the spherically symmetric solution to pure
Gauss-Bonnet field equations is \textquotedblleft extended" to $D=3$
\cite{BHS}. The thermodynamics of the black holes found here, also reveals
some similarities with the ones of pure BHT, where the specific heat is
positive (and linear in the temperature), implying the thermal stability of
these black holes. Naturally, due to the remarkable resemblence between the
BHT new massive gravity in three dimensions and our theory in five dimensions,
one can't help but wonder if there exists suitable generalizations to
arbitrary higher order. We show, in this section, that the answer is
affirmative by presenting a recipe to construct generalizations of our theory
to arbitrary higher order. In Appendix B, we give explicit results for the
quartic case.

First, let us recall how the quadratic invariant $K:=4R^{ab}R_{ab}-\frac
{D}{(D-1)}R^{2}$ can be constructed, which in $D=3$ serves as the Lagrangian
for the BHT new massive gravity. The key step here is to realize the following
identity in arbitrary dimensions.
\begin{equation}
C^{abcd}C_{abcd}=\mathcal{E}_{4}+\left(  \frac{D-3}{D-2}\right)  \left(
4R^{ab}R_{ab}-\frac{D}{(D-1)}R^{2}\right)  .
\end{equation}
which can be rewritten as
\begin{equation}
4R^{ab}R_{ab}-\frac{D}{D-1}R^{2}=\left(  \frac{D-2}{D-3}\right)  \left[
C^{abcd}C_{abcd}-\mathcal{E}_{4}\right]  \label{identityone}%
\end{equation}
where $\mathcal{E}_{4}:=R^{2}-4R_{ab}R^{ab}+R_{abcd}R^{abcd}$ is the four
dimensional Euler density and $C_{abcd}$ is the Weyl tensor. At first sight,
it seems that in three dimensions, the right hand side takes a $0/0$ form
since both the Weyl tensor and the four-dimensional Euler density vanishes
identically in $D=3$. However, if one expands the right hand side in terms of
the Riemann curvature tensor then it factorizes by $(D-3)$ which cancels the
one in the denominator of the preceeding factor and we are left with the
combination on the left hand side. Note that the difference of the quadratic
Weyl invariant and the four-dimensional Euler density can also be written as
\begin{equation}
{\frac{1}{2^{2}}}\delta_{c_{1}d_{1}c_{2}d_{2}}^{a_{1}b_{1}a_{2}b_{2}}\left(
C_{a_{1}b_{1}}^{c_{1}d_{1}}C_{a_{2}b_{2}}^{c_{2}d_{2}}-R_{a_{1}b_{1}}%
^{c_{1}d_{1}}R_{a_{2}b_{2}}^{c_{2}d_{2}}\right)
\end{equation}
where $\delta_{\cdots}^{\cdots}$ is the totally atisymmetric tensor.

Now, we are ready to generalize the identity (\ref{identityone}) for higher
oder. First, consider the following invariant of order $k$
\begin{equation}
{\frac{1}{2^{k}}}\delta_{c_{1}d_{1}\cdots c_{k}d_{k}}^{a_{1}b_{1}\cdots
a_{k}b_{k}}\left(  C_{a_{1}b_{1}}^{c_{1}d_{1}}\cdots C_{a_{k}b_{k}}%
^{c_{k}d_{k}}-R_{a_{1}b_{1}}^{c_{1}d_{1}}\cdots R_{a_{k}b_{k}}^{c_{k}d_{k}%
}\right)  \label{termone}%
\end{equation}
Obviously, the above invariant vanishes in dimensions lower than $2k$.
However, if one expands the Weyl tensor in terms of the Riemann tensor, then
it can be factorized by $(D-2k+1)$. This can be seen as follows. Consider the
basis set of $k$-th order Riemann invariants in arbitrary dimensions. In
$D=2k-1$, not all elements of this set are linearly independent. In fact, the
basis set contains one less invariant than in $D\geq2k$. This is beacuse of
the vanishing of the $k$-th order Lovelock density. Now, after the expanding
in terms of the Riemann tensors, the term (\ref{termone}) will not contain any
$(Riemann)^{k}$. So, this invariant cannot vanish identically in $D=2k-1$
unless it is factorized by $(D-2k+1)$.\footnote{This argument cannot be
extended to dimensions $2k-2$ since one obtains another identity involving the
Riemann invariants which is obtained by contracting the Ricci tensor with the
$(k-1)$-th order Lovelock equation.},\footnote{Further expanding all the Weyl
tensors, one can convince one self that the dimensional dependence of the
coeficient of the term with $k-1$ Riemann tensors and one Ricci tensor must be
$\left(  D-2k+1\right)  /\left(  D-2\right)  $.} We can now divide this factor
out to get a non-vanishing invariant in $D=2k-1$. Thus, we write the $k$th
order generalization of $K$ by evaluating
\begin{equation}
{\frac{1}{2^{k}}}\left(  \frac{1}{D-2k+1}\right)  \delta_{c_{1}d_{1}\cdots
c_{k}d_{k}}^{a_{1}b_{1}\cdots a_{k}b_{k}}\left(  C_{a_{1}b_{1}}^{c_{1}d_{1}%
}\cdots C_{a_{k}b_{k}}^{c_{k}d_{k}}-R_{a_{1}b_{1}}^{c_{1}d_{1}}\cdots
R_{a_{k}b_{k}}^{c_{k}d_{k}}\right)
\end{equation}
in $D=2k-1$. Now, by construction, the trace of the field equation arising
from the above invariant is of second order in all dimensions. Moreover, for
static spherically symmetric spacetimes this invariant must be a sum of the
$k$th Lovelock invariant and a term proportional to $(Weyl)^{k}$. This is
because for static spherically symmetric spacetimes, all the Weyl invariants
are proportional to each other \cite{deserspherical}. Thus, if one subtracts
an appropriate multiple of $(Weyl)^{k}$ from this invariant, then all the
components of the field equation, arising from the resulting invariant will be
of second order. Since, for $k\geq4$, there are more than one linearly
independent Weyl invariants in dimension $2k-1$, one can do this in several
ways. One convenient choice is
\begin{equation}
\mathcal{L}:={\frac{1}{2^{k}}}\left(  \frac{1}{D-2k+1}\right)  \delta
_{c_{1}d_{1}\cdots c_{k}d_{k}}^{a_{1}b_{1}\cdots a_{k}b_{k}}\left(
C_{a_{1}b_{1}}^{c_{1}d_{1}}\cdots C_{a_{k}b_{k}}^{c_{k}d_{k}}-R_{a_{1}b_{1}%
}^{c_{1}d_{1}}\cdots R_{a_{k}b_{k}}^{c_{k}d_{k}}\right)  -\alpha_{k}%
C_{a_{1}b_{1}}^{a_{k}b_{k}}C_{a_{2}b_{2}}^{a_{1}b_{1}}\cdots C_{a_{k}b_{k}%
}^{a_{k-1}b_{k-1}} \label{Lk}%
\end{equation}
where
\begin{equation}
\alpha_{k}={\frac{(D-4)!}{(D-2k+1)!}}{\frac
{[k(k-2)D(D-3)+k(k+1)(D-3)+(D-2k)(D-2k-1)]}{[(D-3)^{k-1}(D-2)^{k-1}%
+2^{k-1}-2(3-D)^{k-1}]}}%
\end{equation}
Again, note that the above invariant vanishes identically in $D\leq2k-2$,
whereas in dimensions $D\geq2k$, it can be expressed as a linear combination
of the Weyl invariants and the $2k$-dimensional Euler density. After replacing
the ansatz%
\begin{equation}
ds_{D}^{2}=-N\left(  r\right)  f\left(  r\right)  dt^{2}+\frac{dr^{2}%
}{f\left(  r\right)  }+r^{2}d\Sigma_{\gamma,2k-3}^{2}%
\end{equation}
in (\ref{Lk}) and carrying out the variation of the action with respect to
$f\left(  r\right)  $ and $N\left(  r\right)  $, one respectively obtains%
\begin{align}
\left(  \gamma-f\right)  ^{k-1}N^{\prime}  &  =0\ ,\\
\left(  \gamma-f\right)  ^{k-1}\left[  \left(  D-2k-1\right)  \left(
\gamma-f\right)  -krf^{\prime}\right]   &  =0\ .
\end{align}
The non-trivial branch of solutions for $D=2k-1$, is given by
\begin{equation}
ds^{2}=-\left(  cr^{2/k}+\gamma\right)  dt^{2}+\frac{dr^{2}}{cr^{2/k}+\gamma
}+r^{2}d\Sigma_{2k-3}^{2}\ , \label{genralizedsol}%
\end{equation}
where $c$ is a integration constant and $d\Sigma_{(2k-3)}$ is the line element
of a Euclidean ($2k-3$)-dimensional space of constant curvature \footnote{In
three dimensions for BHT, $\gamma$ is an integration constant.} $\gamma
=\pm1,0$. For positive $c$ and $\gamma=-1$, this describes a topological black
hole with a horizon located at $r=r_{+}=c^{-\frac{k}{2}}$. The temperature of
the black hole is $1/2\pi kr_{+}$. One can compute the entropy using Wald's
formula and then obtain the mass with respect to the locally flat background
by applying the first law,
\begin{equation}
S\propto\frac{Vol\ (\Sigma_{2k-3})}{r_{+}},\ \ \ \ \ M\propto\frac
{Vol\ (\Sigma_{2k-3})}{r_{+}^{2}}.
\end{equation}

\section{Further comments}

\label{furthercomments}

In this paper we have investigated a new interesting theory of gravity, which
is cubic in curvature. This theory as we have shown, has several remarkable
characteristics such as having second order field equations for static
spherically symmetric ansatz, admittance of Birkhoff's theorem and existence
of a C-function for the black hole solution. This theory is the unique cubic
theory in five dimensions, for which the field equations for the static
spherically symmetric spacetimes are of second order \cite{cubicwork}. As in
the case of Lovelock theory, the admittance of Birkhoff's theorem
\cite{Zegers}, \cite{Deser:2005gr}, suggests the lack of the spin-0 mode in
the linearized theory \cite{Riegert}, \cite{Deser2}. A definite confirmation
to this assertion requires a full Hamiltonian analysis, which can be performed
for example along the lines of Ref. \cite{DERU} , or in a spherically
symmetric minisuperspace approach \cite{Palais}, which is straightforward due
to the second order nature of the theory in this setup. This can be seen from
the Lagrangian, where all the terms that are second order derivatives in the
metric functions are of the form $H(q)$ $\ddot{q}$, which can be integrated by
parts to obtain a first order Lagrangian.

It is natural to expect that, when perturbed around flat space (up to the
leading order), our theory will possess ghost degrees of freedom, since
generic perturbations will break the spherical symmetry and involve fourth
order derivatives. Nevertheless, since this assertion is background dependent,
it would be nice to look for a ghost-free background, in analogy with
Topologically Massive Gravity \cite{Strominger}.

In the nonhomogenous combination, due to the presence of additional scales, it
is natural to expect that the black hole (\ref{solnonhom}) will have different
phases depending on the sign of the specific heat, as is the case for the
black holes studied in \cite{BHS}. It will be also interesting to prove
Birkhoff's theorem for all the higher order generalizations. Work along these
lines is in progress.

It would be interesting to explore the dimensional reduction of the
nonhomogeneous theory to four dimensions, along the lines of Ref.
\cite{DIMRED}.

\textbf{Note}: After the first version of this work was submitted to arXiv,
another paper \cite{MR} appeared where the same cubic invariant in five
dimensions is presented. The authors generalize the cubic invariant in higher
dimensions by requiring second order field equations for static spherically
symmetric spacetimes. However, their generalization in higher dimensions is
nothing but a particular linear combination of the six-dimensional Euler
density and the two linearly independent Weyl invariants. As mentioned above,
since for static spherically symmetric spacetimes, both the Weyl invariants
are proportional to each other, addition of the two invariants with a
particular choice of the relative factor does not contribute to the field
equations. In this case, one obviously obtains the field equations for the
cubic Lovelock theory.

\bigskip

\textit{Acknowledgments.} We thank Andres Anabalon, Eloy Ayon-Beato, Mokhtar
Hassaine, David Kastor, David Tempo, Ricardo Troncoso and Steven Willison, for
useful discussions. This research is partially funded by Fondecyt grants
number 3095018, 11090281, and by the Conicyt grant \textquotedblleft Southern
Theoretical Physics Laboratory\textquotedblright\ ACT-91. The Centro de
Estudios Cient\'{\i}ficos (CECS) is funded by the Chilean Government through
the Millennium Science Initiative and the Centers of Excellence Base Financing
Program of Conicyt. CECS is also supported by a group of private companies
which at present includes Antofagasta Minerals, Arauco, Empresas CMPC, Indura,
Naviera Ultragas and Telef\'{o}nica del Sur. CIN is funded by Conicyt and the
Gobierno Regional de Los R\'{\i}os.

\appendix

\section{General static, spherically symmetric solution for the\newline
non-homogeneous combination}

\bigskip

In this appendix, we consider the general static, spherically symmetric
solution for a generic non-homogeneous combination (\ref{nonhom}). For
simplicity, we restrict to spacetime metrics of the form (\ref{metric1}), with
$f=g$. In this case the components of the field equations $\mathcal{E}%
_{\ t}^{t}$ and $\mathcal{E}_{\ r}^{r}$ are equal and reduce to%

\[
\mathcal{E}_{\ t}^{t}=\mathcal{E}_{\ r}^{r}:=2c_{0}r^{3}-3c_{1}\left(
r^{2}\left(  \gamma-f\right)  \right)  ^{\prime}-6c_{2}\left(  \left(
\gamma-f\right)  ^{2}\right)  ^{\prime}+c_{3}\left(  r^{-2}\left(
\gamma-f\right)  ^{3}\right)  ^{\prime}=0\ ,
\]
which, after defining $F:=\gamma-f$ reduces to%
\[
2c_{0}r^{3}-3c_{1}\left(  r^{2}F\right)  ^{\prime}-6c_{2}\left(  F^{2}\right)
^{\prime}+c_{3}\left(  r^{-2}F^{3}\right)  ^{\prime}=0\ .
\]
This equation can be trivially integrated to obtain the following algebraic
equation
\begin{equation}
c_{0}r^{6}-6c_{1}r^{4}F-12r^{2}c_{2}F^{2}+c_{3}F^{3}+2\mu r^{2}=0\ ,
\label{cubicalgebraic}%
\end{equation}
$\mu$ being the integration constant. The field equations with indices along
the manifold $\Sigma_{3}$, i.e. $\mathcal{E}_{\ j}^{i}$ are given by%
\begin{equation}
\mathcal{E}_{\ j}^{i}=\frac{1}{3r^{2}}\left(  r^{3}\mathcal{E}_{\ r}%
^{r}\right)  ^{\prime}\ \delta_{j}^{i}\ . \label{alongbase}%
\end{equation}
Therefore, the solutions to equation (\ref{cubicalgebraic}), trivially solve
equation (\ref{alongbase}). This again is analogous to Lovelock theories, in
which the problem reduces to solving an algebraic equation, given by the
Wheeler's polynomial \cite{JTW}. Generically there are three branches, which
are asymptotically locally a spacetime of constant curvature $\lambda_{1}$,
$\lambda_{2}$ or $\lambda_{3}$, when the equation (\ref{cubicalgebraic}) has
real roots.

Now, let us examine a particular case for which the solution takes a simple
form. This is the case when the pure cubic theory is appended by a
cosmological term, i.e. $c_{1}=0$ and $c_{2}=0$ in (\ref{nonhom}). This is the
simplest case which admits an asymptotically locally AdS solution. In this
case, the cubic equation reduces to%
\begin{equation}
c_{0}r^{6}+c_{3}F^{3}+2\mu r^{2}=0\ ,
\end{equation}
which is solved by%
\begin{equation}
f\left(  r\right)  =\gamma+\frac{r^{2}}{l^{2}}\left(  1-\frac{3l^{2}\tilde
{\mu}}{r^{4}}\right)  ^{1/3}\ , \label{cailike}%
\end{equation}
where we have defined the AdS radius as $l^{2}:=\left(  c_{3}/c_{0}\right)
^{1/3}$ (chosen to be positive) and the integration constant $\mu$ has been
replaced by $\tilde{\mu}=-2\mu l^{4}/\left(  3c_{3}\right)  $. The spacetime
described by (\ref{cailike}) is asymptotically locally AdS, and describes a
topological black hole for a certain range of the parameters. Note that
expanding around infinity, the subleading term goes as a
Schwarzschild-Tangherlini term ($\tilde{\mu}/r^{2}$), suggesting $\tilde{\mu}$
as the mass parameter.

\section{A new quartic theory of gravity in seven dimensions}

Here, we present the generalization of our theory to quartic Lagrangians in
seven dimensions. Consider the following basis of quartic invariants
\cite{Fulling}.
\begin{align}
&  L_{1}=R^{pqbs}R_{p\ b}^{\ a\ u}R_{a\ q}^{\ v\ w}R_{uvsw},\ L_{2}%
=R^{pqbs}R_{p\ b}^{\ a\ u}R_{a\ u}^{\ v\ w}R_{qvsw},\ L_{3}=R^{pqbs}%
R_{pq}^{\ \ au}R_{b\ a}^{\ v\ w}R_{svuw},\nonumber\\
&  L_{4}=R^{pqbs}R_{pq}^{\ \ au}R_{ba}^{\ \ vw}R_{suvw},\ L_{5}=R^{pqbs}%
R_{pq}^{\ \ au}R_{au}^{\ \ vw}R_{bsvw},\ L_{6}=R^{pqbs}R_{pqb}^{\ \ \ a}%
R_{\ \ \ s}^{uvw}R_{uvwa},\nonumber\\
&  L_{7}=\left(  R^{pqbs}R_{pqbs}\right)  ^{2},\ L_{8}=R^{pq}R^{bsau}%
R_{b\ ap}^{\ v}R_{svuq},\ L_{9}=R^{pq}R^{bsau}R_{bs\ p}^{\ \ v}R_{auvq}%
,\nonumber\\
&  L_{10}=R^{pq}R_{p\ q}^{\ b\ s}R_{\ \ \ b}^{auv}R_{auvs},\ L_{11}%
=RR^{pqbs}R_{p\ b}^{\ a\ \ u}R_{qasu},\ L_{12}=RR^{pqbs}R_{pq}^{\ \ au}%
R_{bsau},\nonumber\\
&  L_{13}=R^{pq}R^{bs}R_{\ p\ b}^{a\ u}R_{aqus},\ L_{14}=R^{pq}R^{bs}%
R_{\ p\ q}^{a\ u}R_{abus},\ L_{15}=R^{pq}R^{bs}R_{\ \ pb}^{au}R_{auqs}%
,\nonumber\\
&  L_{16}=R^{pq}R_{p}^{\ b}R_{\ \ \ q}^{sau}R_{saub},\ L_{17}=R^{pq}%
R_{pq}R^{bsau}R_{bsau},\ L_{18}=RR^{pq}R_{\ \ \ p}^{bsa}R_{bsaq},\nonumber\\
&  L_{19}=R^{2}R^{pqbs}R_{pqbs},\ L_{20}=R^{pq}R^{bs}R_{b}^{\ a}%
R_{psqa},\ L_{21}=RR^{pq}R^{bs}R_{pbqs},\nonumber\\
&  L_{22}=R^{pq}R_{p}^{\ b}R_{q}^{\ s}R_{bs},\ L_{23}=\left(  R^{pq}%
R_{pq}\right)  ^{2},\ L_{24}=RR^{pq}R_{p}^{\ b}R_{qb},\ L_{25}=R^{2}%
R^{pq}R_{pq},\ L_{26}=R^{4}. \label{quarticinvariants}%
\end{align}
In seven dimensions, they are not linearly independent, since they are related
to the eight-dimensional Euler density
\begin{align*}
E_{8}  &  :=-96L_{1}+48L_{2}-96L_{3}+24L_{4}+18L_{5}-48L_{6}+3L_{7}%
+384L_{8}-192L_{9}+192L_{10}-32L_{11}+16L_{12}\\
&  +192L_{13}-192L_{14}+96L_{15}+192L_{16}-24L_{17}-96L_{18}+6L_{19}%
-384L_{20}+96L_{21}-96L_{22}+48L_{23}\\
&  +64L_{24}-24L_{ 25}+L_{26}.
\end{align*}
which vanishes identically in dimensions lower than eight. Now, we consider a
Lagrangian, constructed by taking a linear combination of all the invariants
from the set (\ref{quarticinvariants}), with the following choice of
coefficients
\begin{align}
L^{(4)}:=  &  16 L_{1}+38 L_{2}+\frac{41}{2}L_{4}-14 L_{6}-\frac{141}{16}%
L_{7}+16 L_{8}-32 L_{9}+2 L_{12}-24 L_{14}+8 L_{15}+16 L_{16}+\frac{153}%
{10}L_{17}\nonumber\\
&  -\frac{61}{40}L_{19}+\frac{8}{5}L_{22}-\frac{153}{25}L_{23}-\frac{16}%
{5}L_{24}+\frac{121}{50}L_{25}-\frac{57}{400}L_{26}.
\label{specialquarticinvariant}%
\end{align}
In analogy with the quadratic (BHT new massive gravity) and the cubic case,
the trace of the field equations obtained from the above Lagrangian, being
proportional to the Lagrangian itself, is of second order. In fact, there are
seven Weyl invariants, constructed by taking complete contractions of four
Weyl tensors, which also has this property in arbitrary dimensions. In terms
of the basis of invariants (\ref{quarticinvariants}), they are given as
\begin{align}
W_{1}\text{:=}  &  L_{1}-\frac{8}{5} \left(  L_{8}-\frac{L_{9}}{4}\right)
-\frac{4 L_{10}}{5}+\frac{2}{15} \left(  L_{11}-\frac{L_{12}}{4}\right)
-\frac{22 L_{13}}{25}+\frac{16 L_{14}}{25}-\frac{4 L_{15}}{25}-\frac{12
L_{16}}{25}+\frac{2 L_{17}}{25}\nonumber\\
&  +\frac{8 L_{18}}{25}-\frac{L_{19}}{50}+\frac{172 L_{20}}{125}-\frac{42
L_{21}}{125}+\frac{201 L_{22}}{625}-\frac{103 L_{23}}{625}-\frac{416 L_{24}%
}{1875}+\frac{52 L_{25}}{625}-\frac{13 L_{26}}{3750}\nonumber\\
W_{2}\text{:=}  &  L_{2}+\frac{8}{5} \left(  L_{8}-\frac{L_{9}}{4}\right)
-\frac{2}{15} \left(  L_{11}-\frac{L_{12}}{4}\right)  +\frac{12 L_{13}}%
{25}-\frac{28 L_{14}}{25}+\frac{4 L_{15}}{25}+\frac{8 L_{16}}{25}-\frac{4
L_{18}}{25}+\frac{L_{19}}{150}\nonumber\\
&  -\frac{136 L_{20}}{125}+\frac{46 L_{21}}{125}-\frac{138 L_{22}}{625}%
+\frac{104 L_{23}}{625}+\frac{308 L_{24}}{1875}-\frac{148 L_{25}}{1875}%
+\frac{37 L_{26}}{11250}\nonumber\\
W_{3}\text{:=}  &  L_{3}-\frac{8 L_{8}}{5}+\frac{2 L_{9}}{5}-\frac{2 L_{10}%
}{5}+\frac{2 L_{11}}{15}-\frac{L_{12}}{30}+\frac{12 L_{14}}{25}-\frac{12}{25}
\left(  L_{13}-\frac{L_{15}}{2}\right)  -\frac{3 L_{15}}{5}-\frac{2 L_{16}}%
{5}+\frac{L_{17}}{25}\nonumber\\
&  +\frac{6 L_{18}}{25}-\frac{L_{19}}{75}+\frac{144 L_{20}}{125}-\frac{34
L_{21}}{125}+\frac{152 L_{22}}{625}-\frac{76 L_{23}}{625}-\frac{332 L_{24}%
}{1875}+\frac{122 L_{25}}{1875}-\frac{61 L_{26}}{22500}\nonumber\\
W_{4}\text{:=}  &  L_{4}-\frac{8 L_{9}}{5}+\frac{2 L_{12}}{15}+\frac{16}{25}
\left(  L_{13}-\frac{L_{15}}{2}\right)  +\frac{24 L_{15}}{25}+\frac{16 L_{16}%
}{25}-\frac{8 L_{18}}{25}+\frac{L_{19}}{75}-\frac{96 L_{20}}{125}+\frac{16
L_{21}}{125}\nonumber\\
&  -\frac{168 L_{22}}{625}+\frac{24 L_{23}}{625}+\frac{96 L_{24}}{625}%
-\frac{68 L_{25}}{1875}+\frac{17 L_{26}}{11250}\nonumber\\
W_{5}\text{:=}  &  L_{5}-\frac{16 L_{9}}{5}+\frac{4 L_{12}}{15}+\frac{32
L_{13}}{25}+\frac{32 L_{15}}{25}+\frac{32 L_{16}}{25}-\frac{16 L_{18}}%
{25}+\frac{2 L_{19}}{75}-\frac{192 L_{20}}{125}+\frac{32 L_{21}}{125}%
-\frac{336 L_{22}}{625}\nonumber\\
&  +\frac{48 L_{23}}{625}+\frac{192 L_{24}}{625}-\frac{136 L_{25}}{1875}%
+\frac{17 L_{26}}{5625}\nonumber\\
W_{6}\text{:=}  &  L_{6}-\frac{8 L_{10}}{5}+\frac{16 L_{14}}{25}-\frac{28
L_{16}}{25}+\frac{4 L_{17}}{25}+\frac{8 L_{18}}{25}-\frac{2 L_{19}}{75}%
+\frac{112 L_{20}}{125}-\frac{32 L_{21}}{125}+\frac{196 L_{22}}{625}-\frac{108
L_{23}}{625}\nonumber\\
&  -\frac{112 L_{24}}{625}+\frac{136 L_{25}}{1875}-\frac{17 L_{26}}%
{5625}\nonumber\\
W_{7}\text{:=}  &  L_{7}-\frac{8 L_{17}}{5}+\frac{2 L_{19}}{15}+\frac{16
L_{23}}{25}-\frac{8 L_{25}}{75}+\frac{L_{26}}{225}.
\label{quarticweylinvariants}%
\end{align}

Note that the Weyl invariants satisfy the following identity
\begin{equation}
E_{8}:=-96W_{1}+48W_{2}-96W_{3}+24W_{4}+18W_{5}-48W_{6}+3W_{7}\equiv0.
\end{equation}
This implies that only six of the above $W_{i}$'s are linearly independent in
seven dimensions. However, the invariant (\ref{specialquarticinvariant}) is
linearly independent of the set of Weyl invariants
(\ref{quarticweylinvariants}) and consequently does not transform covariantly
under Weyl rescalings.

Let us consider the following action
\begin{equation}
I_{4}=\kappa_{4}\int\sqrt{-g} L^{(4)} d^{7}x. \label{actionquartic}%
\end{equation}
Now, consider a static spherically symmetric spacetime described by the line
element
\begin{equation}
ds^{2}=-N(r)f\left(  r\right)  dt^{2}+\frac{dr^{2}}{f\left(  r\right)  }%
+r^{2}d\Sigma_{5,\gamma}^{2}\ , \label{sevendimmetric}%
\end{equation}
where $d\Sigma_{5,\gamma}$ is the line element of an Euclidean
five-dimensional space of constant curvature $\gamma=\pm1,0$. Using the
minisuperspace method \cite{Palais}, we obtain the following two second order
equations for $f(r)$ and $N(r)$:
\begin{align}
N^{\prime3}  &  =0 ,\\
(\gamma-f(r))^{3}[2rf^{\prime}(r)+\gamma-f(r)])  &  =0 .
\end{align}
The nontrivial solution for the pure quartic case is given by:%
\[
ds^{2}=-\left(  cr^{1/2}+\gamma\right)  dt^{2}+\frac{dr^{2}}{cr^{1/2}+\gamma
}+r^{2}d\Sigma_{5,\gamma}^{2}\ ,
\]
where $c$ is an integration constant. As in the pure cubic theory, for the
topological case $\gamma=-1$ and positive $c$, this metric describes an
asymptotically locally flat black hole with horizon radius $r=r_{+}:=1/c^{2}$,
that can be rewritten as%
\begin{equation}
ds^{2}=-\left(  \left(  \frac{r}{r_{+}}\right)  ^{1/2}-1\right)  dt^{2}%
+\frac{dr^{2}}{\left(  \frac{r}{r_{+}}\right)  ^{1/2}-1}+r^{2}d\Sigma
_{5,\gamma}^{2}\ .
\end{equation}

In this case, the horizon is described by a quotient of the five-dimensional
hyperbolic space $H_{5}/\Gamma$, where $\Gamma$ is a freely acting discrete
subgroup of $O\left(  5,1\right)  $.

The temperature of the black hole is%
\[
T=\frac{1}{8\pi}\frac{1}{r_{+}}\ .
\]
Using Wald's formula for the entropy and then the first law to compute the
mass, we obtain%
\begin{align}
S  &  =\frac{480\pi}{r_{+}}\kappa_{4}Vol\left(  \Sigma_{5}\right) \\
M  &  =\frac{30Vol\left(  \Sigma_{5}\right)  \kappa_{4}}{r_{+}^{2}}%
\end{align}
respectively. Finally, the specific heat of this black hole is given by%
\begin{equation}
C=3840\pi^{2}\kappa_{4}T\ Vol\left(  \Sigma_{5}\right)  \ ,
\end{equation}
which, being positive, implies that the black hole is thermally stable. Note
that the functional dependence of all the expressions remain the same as their
five-dimensional, cubic counterpart.

Let us further consider a generic linear combination of the quartic and all
possible lower order Lovelock terms:
\begin{equation}
I_{7}=\kappa_{4}\int\sqrt{-g}\sum_{i=0}^{4}\alpha_{i}L^{(i)}d^{7}x,
\label{actionquarticnh}%
\end{equation}
where $L^{0}:=1$, $L^{1}:=R$ and
\begin{align}
L^{2}:=  &  R_{abcd}R^{abcd}-4R_{ab}R{ab}+R^{2},\\
L^{3}:=  &  2R^{abcd}R_{cdef}R_{\ \ ab}^{ef}+8R_{\ \ cd}^{ab}R_{\ \ bf}%
^{ce}R_{\ \ ae}^{df}+24R^{abcd}R_{cdbe}R_{\ a}^{e}+\\
&  3RR^{abcd}R_{abcd}+24R^{abcd}R_{ac}R_{bd}+16R^{ab}R_{bc}R_{\ a}%
^{c}-12RR^{ab}R_{ab}+R^{3}.
\end{align}
The field equations for spherically symmetric ansatz (\ref{sevendimmetric})
reduce to%
\begin{equation}
-r^{5}\alpha_{0}+5\left(  r^{4}F\right)  ^{\prime}\alpha_{1}-60\left(
r^{2}F^{2}\right)  ^{\prime}\alpha_{2}+120\left(  F^{3}\right)  ^{\prime
}\alpha_{3}-30\left(  r^{-2}F^{4}\right)  ^{\prime}\alpha_{4}=0\ ,
\end{equation}
where $F=\gamma-f$. The above equation can be trivially integrated as%
\[
-r^{8}\alpha_{0}+30r^{6}F\alpha_{1}-360r^{4}F^{2}\alpha_{2}+720F^{3}%
r^{2}\alpha_{3}-180F^{4}\alpha_{4}+6r^{2}\mu=0\ .
\]
Here $\mu$ is an integration constant. Obviously, depending on the roots of
the above polynomial equation, the metric may then describe a black hole.

The maximally symmetric (see (\ref{constantcurvature})) solutions of this
theory fulfill%
\begin{equation}
\alpha_{0}+30\lambda\alpha_{1}+360\lambda^{2}\alpha_{2}+720\lambda^{3}%
\alpha_{3}+180\lambda^{4}\alpha_{4}=0\ . \label{maximalequation}%
\end{equation}
Generically, a spherically symmetric solution will asymptotically approach a
maximally symmetric background of constant curvature $\lambda_{i}$, where
$\lambda_{i}$ is a real root of (\ref{maximalequation}). When the coupling
constants $\alpha_{i}$ are such that%
\begin{equation}
\alpha_{0}=180\frac{\alpha_{3}^{4}}{\alpha_{4}^{3}},\ \alpha_{1}%
=24\frac{\alpha_{3}^{3}}{\alpha_{4}^{2}},\ \text{and\ }\alpha_{2}%
=3\frac{\alpha_{3}^{2}}{\alpha_{4}}\ ,
\end{equation}
equation (\ref{maximalequation}) reduces to%
\begin{equation}
\frac{\left(  \alpha_{4}+\alpha_{3}\lambda\right)  ^{4}}{\alpha_{4}^{3}}=0\ ,
\end{equation}
and the four maximally symmetric vaccua of the theory coincide. The
spherically symmetric solution is then given by
\begin{equation}
ds^{2}=-\left(  \frac{r^{2}}{l^{2}}-\tilde{\mu}r^{1/2}+\gamma\right)
dt^{2}+\frac{dr^{2}}{\frac{r^{2}}{l^{2}}-\tilde{\mu}r^{1/2}+\gamma}%
+r^{2}d\Sigma_{5,\gamma}^{2}\ ,
\end{equation}
where $l^{2}:=\alpha_{3}/\alpha_{4}$ is the square of the AdS radius and
$\tilde{\mu}$ is the rescaled integration constant. Depending on the values of
the parameters, the metric may then describe an asymptotically AdS black hole
or a topological black hole. The fall-off at infinity is slower than the one
in GR and the spectrum of spacetimes is the same as in the cubic case, which
we described at the end of section (\ref{nonhomogenouscombination}). However,
as mentioned earlier, for static spherically symmetric spacetimes, all the
Weyl invariants are proportional to each other. Specifically, in seven
dimensions they are related by
\begin{equation}
\frac{W_{1}}{93}=\frac{W_{2}}{191}=\frac{W_{3}}{11}=\frac{W_{4}}{226}%
=\frac{W_{5}}{452}=\frac{W_{6}}{328}=\frac{W_{7}}{1000}.
\end{equation}
This means that one can always add an arbitrary combination of the Weyl
invariants $\sum_{i=1}^{7}C_{i}W_{i}$ to the Lagrangian
(\ref{specialquarticinvariant}), such that
\begin{equation}
93C_{1}+191C_{2}+11C_{3}+226C_{4}+452C_{5}+328C_{6}+1000C_{7}=0.
\end{equation}
without affecting the field equations and their spherically symmetric solutions.


\begin{thebibliography}{99}                                                                                               %


\bibitem {LOVELOCK}D.~Lovelock,
J.\ Math.\ Phys.\ \textbf{12}, 498 (1971).


\bibitem {BOULWARE-DESER}D.~G.~Boulware and S.~Deser,
Phys.\ Rev.\ Lett.\ \textbf{55}, 2656 (1985).


\bibitem {BHT1}E.~A.~Bergshoeff, O.~Hohm and P.~K.~Townsend,
Phys.\ Rev.\ Lett.\ \textbf{102}, 201301 (2009) [arXiv:0901.1766 [hep-th]].


\bibitem {NO}M.~Nakasone and I.~Oda,
Prog.\ Theor.\ Phys.\ \textbf{121}, 1389 (2009) [arXiv:0902.3531 [hep-th]].


\bibitem {far}M.~Farhoudi,
Int.\ J.\ Mod.\ Phys.\ D \textbf{14}, 1233 (2005) [arXiv:gr-qc/9511047].


\bibitem {cubicwork}J.~Oliva and S.~Ray,
arXiv:1004.0737 [gr-qc].


\bibitem {Charmousis}C.~Charmousis and J.~F.~Dufaux,
Class.\ Quant.\ Grav.\ \textbf{19}, 4671 (2002) [arXiv:hep-th/0202107].


\bibitem {Aliev:2007dp}A.~N.~Aliev, H.~Cebeci and T.~Dereli,
Class.\ Quant.\ Grav.\ \textbf{24}, 3425 (2007) [arXiv:gr-qc/0703011].


\bibitem {Zegers}R.~Zegers,
J.\ Math.\ Phys.\ \textbf{46}, 072502 (2005) [arXiv:gr-qc/0505016].


\bibitem {Deser:2005gr}S.~Deser and J.~Franklin,
Class.\ Quant.\ Grav.\ \textbf{22}, L103 (2005) [arXiv:gr-qc/0506014].


\bibitem {DOT2}G.~Dotti, J.~Oliva and R.~Troncoso,
Phys.\ Rev.\ D \textbf{76}, 064038 (2007) [arXiv:0706.1830 [hep-th]].


\bibitem {DOT3}G.~Dotti, J.~Oliva and R.~Troncoso,
Int.\ J.\ Mod.\ Phys.\ A \textbf{24}, 1690 (2009) [arXiv:0809.4378 [hep-th]].


\bibitem {Zam}A.~B.~Zamolodchikov,
JETP Lett.\ \textbf{43}, 730 (1986) [Pisma
Zh.\ Eksp.\ Teor.\ Fiz.\ \textbf{43}, 565 (1986)].


\bibitem {sa}V.~Sahakian,
Phys.\ Rev.\ D \textbf{62}, 126011 (2000) [arXiv:hep-th/9910099].


\bibitem {golds}K.~Goldstein, R.~P.~Jena, G.~Mandal and S.~P.~Trivedi,
JHEP \textbf{0602}, 053 (2006) [arXiv:hep-th/0512138].


\bibitem {dimitru}D.~Astefanesei, H.~Nastase, H.~Yavartanoo and S.~Yun,
JHEP \textbf{0804}, 074 (2008) [arXiv:0711.0036 [hep-th]].


\bibitem {KASTORC}M.~M.~Anber and D.~Kastor,
JHEP \textbf{0805}, 061 (2008) [arXiv:0802.1290 [hep-th]].


\bibitem {sinha}A.~Sinha,
arXiv:1003.0683 [hep-th].


\bibitem {Wald}R.~M.~Wald,
Phys.\ Rev.\ D \textbf{48}, 3427 (1993) [arXiv:gr-qc/9307038].


\bibitem {JM}T.~Jacobson, G.~Kang and R.~C.~Myers,
Phys.\ Rev.\ D \textbf{52}, 3518 (1995) [arXiv:gr-qc/9503020].


\bibitem {DOTWorm}G.~Dotti, J.~Oliva and R.~Troncoso,
Phys.\ Rev.\ D \textbf{75}, 024002 (2007).


\bibitem {COT}D.~H.~Correa, J.~Oliva and R.~Troncoso,
J. High Energy Phys. \textbf{0808}, 081 (2008).


\bibitem {CamelloMarameo2}F.~Canfora and A.~Giacomini, \emph{Vacuum static
compactified wormholes in eight-dimensional Lovelock theory,} arXiv:0808.1597
[hep-th].


\bibitem {Cai}R.~G.~Cai and K.~S.~Soh,
Phys.\ Rev.\ D \textbf{59}, 044013 (1999).


\bibitem {BHStopo}R.~Aros, R.~Troncoso and J.~Zanelli,
Phys.\ Rev.\ D \textbf{63}, 084015 (2001).


\bibitem {GOT}G.~Giribet, J.~Oliva and R.~Troncoso,
J. High Energy Phys. \textbf{0605}, 007 (2006).


\bibitem {KM}D.~Kastor and R.~B.~Mann,
J. High Energy Phys. \textbf{0604}, 048 (2006).


\bibitem {ROT}A.~Anabalon, N.~Deruelle, Y.~Morisawa, J.~Oliva, M.~Sasaki,
D.~Tempo and R.~Troncoso,
Class.\ Quant.\ Grav.\ \textbf{26}, 065002 (2009) [arXiv:0812.3194 [hep-th]].


\bibitem {topblack}R.~B.~Mann,
Class.\ Quant.\ Grav.\ \textbf{14}, L109 (1997) [arXiv:gr-qc/9607071].


\bibitem {HenTai}M.~Henneaux and C.~Teitelboim,
Commun.\ Math.\ Phys.\ \textbf{98}, 391 (1985).


\bibitem {BHS}J.~Crisostomo, R.~Troncoso and J.~Zanelli,
Phys.\ Rev.\ D \textbf{62}, 084013 (2000) [arXiv:hep-th/0003271].


\bibitem {HIDEKI}H.~Maeda,
Phys.\ Rev.\ D \textbf{78}, 041503 (2008) [arXiv:0805.4025 [hep-th]].


\bibitem {Riegert}R. Bach Math. Z. 9, 110 (1921); R.~J.~Riegert,
Phys.\ Rev.\ Lett.\ \textbf{53}, 315 (1984)


\bibitem {Deser2}S.~Deser and B.~Tekin,
Class.\ Quant.\ Grav.\ \textbf{20}, 4877 (2003) [arXiv:gr-qc/0306114].


\bibitem {DERU}N.~Deruelle, M.~Sasaki, Y.~Sendouda and D.~Yamauchi,
arXiv:0908.0679 [hep-th].


\bibitem {Palais}R. Palais. Comm. Math. Phys. Volume 69, Number 1 (1979), 19-30.

\bibitem {Strominger}D.~Anninos, W.~Li, M.~Padi, W.~Song and A.~Strominger,
JHEP \textbf{0903}, 130 (2009) [arXiv:0807.3040 [hep-th]].


\bibitem {ATZ}R.~Aros, R.~Troncoso and J.~Zanelli,
Phys.\ Rev.\ D \textbf{63}, 084015 (2001) [arXiv:hep-th/0011097].


\bibitem {Tempo}J.~Oliva, D.~Tempo and R.~Troncoso,
JHEP \textbf{0907}, 011 (2009) [arXiv:0905.1545 [hep-th]].


\bibitem {AGH}E.~Ayon-Beato, A.~Garbarz, G.~Giribet and M.~Hassaine,
Phys.\ Rev.\ D \textbf{80}, 104029 (2009) [arXiv:0909.1347 [hep-th]].


\bibitem {BHT2}E.~A.~Bergshoeff, O.~Hohm and P.~K.~Townsend,
Phys.\ Rev.\ D \textbf{79}, 124042 (2009) [arXiv:0905.1259 [hep-th]].


\bibitem {DIMRED}F.~Canfora, A.~Giacomini, R.~Troncoso and S.~Willison,
Phys.\ Rev.\ D \textbf{80}, 044029 (2009) [arXiv:0812.4311 [hep-th]].


\bibitem {deserspherical}S.~Deser and A.~V.~Ryzhov,
Class.\ Quant.\ Grav.\ \textbf{22}, 3315 (2005) [arXiv:gr-qc/0505039].


\bibitem {MR}R.~C.~Myers and B.~Robinson,
arXiv:1003.5357 [gr-qc].


\bibitem {JTW}J.~T.~Wheeler,
Nucl.\ Phys.\ B \textbf{273}, 732 (1986).


\bibitem {Fulling}S.~A.~Fulling, R.~C.~King, B.~G.~Wybourne and
C.~J.~Cummins,
Class.\ Quant.\ Grav.\ \textbf{9} (1992) 1151.

\end{thebibliography}
\end{document}